# Insulating phase at low temperature in ultrathin $La_{0.8}Sr_{0.2}MnO_3$ films

Yaqing Feng[1], Kui-juan Jin[1,2*], Lin Gu[1], Xu He[1], Chen Ge[1], Qing-hua Zhang[3], Min He[1], Qin-lin Guo[1], Qian Wan[1], Meng He[1], Hui-bin Lu[1] and Guozhen Yang[1,2].

*1. Institute of Physics, Chinese Academy of Sciences, Beijing 100190, China*

*2. Collaborative Innovation Center of Quantum Matter, Beijing 100190, China*

*3 School of Materials Science and Engineering, State Key Lab of New Ceramics and Fine Processing, Tsinghua University, Beijing 100084, China*

[*]Correspondence and requests for materials should be addressed to Kui-juan Jin (email: kjjin@iphy.ac.cn).

**Abstract**

Metal-insulator transition is observed in the $La_{0.8}Sr_{0.2}MnO_3$ thin films with thickness larger than 5 unit cells. Insulating phase at lower temperature appeared in the ultrathin films with thickness ranging from 6 unit cells to 10 unit cells and it is found that the Mott variable range hopping conduction dominates in this insulating phase at low temperature with a decrease of localization length in thinner films. A deficiency of oxygen content and a resulted decrease of the Mn valence have been observed in the ultrathin films with thickness smaller than or equal to 10 unit cells by studying the aberration-corrected scanning transmission electron microscopy and electron energy loss spectroscopy of the films. These results suggest that the existence of the oxygen vacancies in thinner films suppresses the double-exchange mechanism and contributes to the enhancement of disorder, leading to a decrease of the Curie temperature and the low temperature insulating phase in the ultrathin films. In addition, the suppression of the magnetic properties in thinner films indicates stronger disorder of magnetic moments, which is considered to be the reason for this decrease of the localization length.

**Introduction**

The interface between two different perovskite oxides can show unexpected and intriguing properties, such as high conductivity or even superconductivity at the interface of two insulating oxides[1], and these achievements have been enabled by the rapid developments in oxide thin film fabrication techniques which can produce sharp interfaces with atomic-scale precision[2-3]. Particularly, the interface between the mixed-valence manganite thin films and other insulating oxides provides magnetic tunnel junctions (MTJ), which can be utilized as the spin injectors or spin valve[4-5]. Therefore, the ultrathin manganite film has been an important ingredient for spintronics and extensive studies into the behavior and mechanism of the manganite thin films are highly required.

Many reports have shown that reducing the thickness results in the suppression of both the magnetic and transport properties of the thin films[6-7]. Moreover, with the thickness of the manganite films dropping down to a certain value, the "dead layer" can be induced, and the manganite films do not exhibit metal phase at very low temperature[8]. Epitaxial strain in the manganite thin films, which can induce structural distortions and produce spin, orbital and charge reconstructions at the interface, has been proposed to play an important role in such behaviors[9-10]. However, understanding for these deviations from the bulk material still remains elusive, and many other factors have been found to account for such behaviors, such as the oxygen vacancies[11-12] and charge transfer at the interfaces[13]. In addition, Liao et al. have demonstrated that the nonmetallic behavior in LSMO thin films is mainly due to the

disorder–induced localization effects which can be amplified by the reduction in dimensionality[14]. Recently, the metal-insulator transition which has never been reported in the bulk $La_{0.9}Sr_{0.1}MnO_3$ has been observed in the ultrathin $La_{0.9}Sr_{0.1}MnO_3$ films[7], indicating that various interesting phenomena still remain to be discovered and understood in the ultrathin manganite films.

In this work, we systematically studied the magnetic and transport properties of the $La_{0.8}Sr_{0.2}MnO_3$ (LSMO) thin films deposited on the $SrTiO_3$ (STO) substrate with thickness varying from 5 unit cells (u.c.) to 50 u.c.. The Curie temperature as well as the magnetic moment of the films decreases when the film becomes thinner. It is found that the LSMO film with thickness varying from 6 u.c. to 10 u.c., not only features a metal-insulator transition at higher temperatures, but becomes insulating again in the low temperature range. We found that this low-temperature insulating behaviors follow the Mott variable range hopping (VRH) law. In addition, a deficiency of the oxygen content was detected in thinner films which results in the decrease of the Mn valence. This result explains the suppression of both the magnetic and transport properties of the ultrathin films and demonstrates that this low temperature insulating phase appeared in ultrathin films is mainly caused by the localization of charge carries due to the existence of enhanced disorder. Meanwhile, though our fitting result indicates that the Mott VRH mechanism dominates in the low temperature insulating phase in ultrathin LSMO films, we cannot completely exclude other mechanisms such as phase separation[15], weak localization over the influence of defects[16] or grain-boundary scattering[16].

**Results**

**Samples preparation and characterization.** LSMO films were deposited on (001) cut $TiO_2$-terminated STO substrates by a laser molecular beam epitaxy (Laser-MBE) system and details on film growth conditions are given in the Methods section. Reflection high-energy electron diffraction (RHEED) oscillations of the 10 u.c. film is recorded during the growth process, as shown in Fig. 1a, and the diffraction pattern of the film (inset of Fig. 1a) indicates a flat surface. The surface topography of the films is further characterized by the atomic force microscopy (AFM) and an atomically flat step-and-terrace surface can be seen in the AFM image of the 10 u.c. film (Fig. 1b).

The crystal structure of the LSMO thin films is characterized by the high-resolution transmission electronic microscopy (TEM), and the high-angle annular dark-field (HAADF) micrograph of the 10 u.c. LSMO film on STO taken by an aberration-corrected STEM is shown in Fig. 1c. In the HAADF micrograph, the brightness is approximately proportional to $Z^2$, so the darker area in Fig. 1c corresponds to the STO substrate, and the brighter one to the 10 u.c. LSMO film. The abrupt contrast between STO and LSMO in Fig.1c shows a well-defined interface. A zoomed and filtered image of the rectangular part in Fig. 1c is presented in Fig. 1d, and the intensity line profiles of the region marked by the red arrow in Fig. 1d is shown in Fig. 1e. Here, the red arrow in Fig. 1d is arranged in a way to cross both the A-site cations (i.e. La/Sr) and the B-site cations (i.e. Mn/Ti). The different contrast of La and Sr (Fig. 1e) provides an accurate localization of the interface indicated by the yellow dashed line in Fig. 1e. It can also be seen that the intensity of the first La/SrO layer deposited on the TiO$_2$-terminated STO substrate is between that of the pure Sr layer in STO and the second La/SrO layer (the 7th peak in Fig.1e) in LSMO, indicating the Sr richness of the first La/SrO layer (the 5th peak in Fig.1e) near the interface. While the Mn/Ti intermixing at the interface is undetectable in Fig. 1e, for the atomic number of Mn and Ti is pretty close and that the contrast of the HAADF micrograph is dominated by the heavy A-site cations.

**Suppression of magnetic properties in ultrathin LSMO films.** Fig. 2a and b show the hysteresis loops (10 K) and the temperature dependence of magnetization (200 Oe, field cooling) of these thin films, respectively. As can be seen in Fig. 2a, the coercive field $H_C$ increases drastically when the thickness of the film is decreased from 30 u.c. to 10 u.c., indicating that the ultrathin films are magnetically significantly harder than the thick films. This might also suggest that both the crystal structure and the domain structure of the ultrathin films are different from that of the thick ones[17], which would be due to the presence of more disorder and deformations since the ultrathin film is more strained compared with the thick ones.

Meanwhile, it can be seen in Fig. 2b that the in-plane magnetization becomes weaker with decreasing temperature below 105 K for comparatively thick films (30 u.c. and 50 u.c.), in contrast to thinner films. We think this phenomenon indicates changes of domain structure in thick LSMO films at 105 K, and we assume this anomaly in M (T) of thick films is induced by the cubic-to-tetragonal phase transition at $T_{CT}$ =105 K in the substrate STO[18-19] since the elastic coupling across interface transmits changes of the substrate into the overgrown films[19-20]. Moreover, at temperatures below 105 K, the presence of stripe domains with magnetic moments tilted up and down from the film plane have been observed in manganite thin films grown on STO substrates which is triggered by the STO transition[21-22]. In this picture, antiferromagnetically coupled regions of antiparallel domains could appear[22], which accounts for the reduction of magnetization at 105 K. Surprisingly, resistivity measurements of the thick films as shown in Fig. 3a do not reveal any anomaly at 105 K, suggesting that the low temperature conductivity of the thick films are not affected much by the phase transition of STO.

However, no anomaly in M (T) at 105 K can be detected in ultrathin films (with thickness smaller than or equal to 10 u.c.). This can be attributed to the presence of small scale defects and strong disorder which distributes throughout the films and thus dominates the low temperature magnetization. Therefore, the effect of STO phase transition doesn't dominate in the magnetic anisotropy of the ultrathin films at 105 K. Analysis of disorder existing in ultrathin films is presented in the discussion part.

The saturation magnetization $M_S$ and the Curie temperature $T_C$ of these films are summarized in Fig. 2c. Here, $M_S$ is defined as the magnetization under the magnetic field of 5 kOe and $T_C$ is defined as the temperature where (dM/dT) reaches the maximum. As can be seen in Fig. 2c, $T_C$ of the 50 u.c. film is 305 K which is comparable to the typical value of about 309 K of the bulk $La_{0.8}Sr_{0.2}MnO_3$[23], and no obvious change was observed in $T_C$ when the thickness of the film decreases to 30 u.c.. When reducing the thickness to 10 u.c., both $M_S$ and $T_C$ of the film become much smaller than that of the bulk LSMO and decrease monotonously when further decreasing the thickness from 10 u.c. to 5 u.c. This suppression of the magnetic properties can be attributed to the weak tensile strain from the STO substrates which induces structural distortion[24] and stretches the Mn-O bonds[25] at or near the LSMO/STO interfaces. The existence of oxygen vacancies is another reason, which will be discussed later in this work. Furthermore, the greatly suppressed magnetic properties of ultrathin films indicate that the magnetic moments are not all aligned in the same direction, which implies presence of disorder of magnetic moments in ultrathin films. This disorder of magnetic moments in ultrathin films is closely related to the substrate-driven strain, as well as the existence of more defects due to the reduced dimensionality and increased inhomogeneity.

**Theoretical analysis of the low-temperature insulating phase appeared in ultrathin LSMO films.** The temperature dependence of resistivity for the investigated thin films with thicknesses varying from 50 u.c. to 5 u.c. is shown in Fig. 3a. When the thickness is 50 u.c. or 30 u.c., the metal-insulator transition appears at room temperature and the resistivity monotonously decreases with decreasing temperature in the low temperature range below the metal-insulator temperature $T_{MI}$, which is very close to that of the bulk LSMO[23]. When the thickness is decreased from 10 u.c. to 6 u.c., the film becomes insulating again in the low temperature range and features a resistivity upturn. Moreover, in the case of the 6 u.c. film, the resistivity is metallic only for a small temperature range below $T_{MI}$. Obviously, $T_{min}$, defined as the temperature at which this resistivity upturn occurs, is dependent on thickness, increasing from 70 K to 165 K when decreasing the thickness from 10 u.c. to 6 u.c.

(Fig. 3d). When the thickness is reduced to 5 u.c., the resistivity of the film increases sharply and no metallic phase can be detected. Moreover, when we plot the resistivity of the 5 u.c. film in Fig. 3c as ln ($\rho$) vs $T^{-1/2}$, a linear dependence was found. This indicates that its transport behavior is consistent with the Efros-Shklovskii (ES) VRH mechanism[26] with $\rho = \rho_0 \exp(T_0/T)^{0.5}$, which means that the strong disorder effect induced localization plays a main role in the transport mechanism, and the distance for an electron to find an energetically suitable site is long.

In order to reveal the origin of the resistivity upturn in the ultrathin films, we have fitted the low-temperature transport properties with the 3D Mott VRH[26] model $\rho = \rho_0 \exp(T_0/T)^{0.25}$ and found that the fitting results, as denoted by the solid lines in Fig. 3a and b, agree well with our experimental data. This result indicates that the resistivity upturn emerges due to the competition of the two conduction mechanism. At temperatures higher than *Tmin*, the double exchange and phase separation mechanisms are the dominant mechanisms in the LSMO structure. As the temperature decreases, the disorder effect becomes dominant and the resistance upturn emerges. When the local potential is strongly disordered, the electrons are in the bounded states and need energy to hop from site to site. At low temperatures, only part of the carriers can be excited and hop between the sites which are within an energy range proportional to the temperature, instead of between nearest neighbors. Since the most probable hopping length $a \approx 0.4\xi_L(T_0/T)^{0.25}$ increases with decreasing temperature[26], the probability of hopping is reduced, which increases the resistance at low temperature.

The parameter $T_0$ is related to DOS at the Fermi level $N(\varepsilon_F)$ and the localization length $\xi_L$. The localization length $\xi_L$, which can be used to describe the spreading of the wave function and how disordered the structure is, are estimated using the Mott VRH law $\xi_L =[21/k_B T_0 N(E_F)]^{\frac{1}{3}}$ [26], with $k_B$ as the Boltzman constant. The density of states at the Fermi level of the bulk material $La_{1-x}Sr_xMnO_3$ at $x=0.2$ of $N(\varepsilon_F)=2.4\times 10^{22}$/eV cm$^3$ are used as an approximation[27]. We can see that $\xi_L$ decreases monotonically with decreasing thickness from Fig. 3d, which is a clear indication that the disorder effect is stronger in thinner films. Here the estimated localization lengths in 6 u.c. and 8 u.c. films are smaller than the lower limit of the distance of the nearest Mn atoms, which may because we use the value of DOS in the bulk. According to Viret, et al[28], the DOS at Fermi energy are smaller when the spins are randomly aligned.

Our results show a crossover from 3D Mott VRH to 2D ES VRH as thickness decreases, which is different from the results of $La_{0.67}Sr_{0.33}MnO_3$ by Liao et al[14]. The critical thickness below which VRH are the dominant mechanism in their samples are smaller (6 u.c. for strained and 3 u.c. for unstrained samples), which are more 2D like. The carrier density in the $La_{0.8}Sr_{0.2}MnO_3$ near the Fermi energy is smaller, resulting in a longer hopping length and more energy needed for the charge to move, which can be the reason for the ES VRH in the films as thin as 5 u.c..

**Discussion**

While the reduction in dimensionality can decrease the localization length because the carriers near the interface can find fewer sites to hop around, the stronger disorder is another reason for the reduced localization lengths in thinner films. To elucidate why, it is essential to find out the possible types of disorder existed in the ultrathin films. In the LSMO thin films, the disorder can be induced by the different A-site cations, namely the La and Sr cations, since they are different both in the valence states and sizes[29]. The orbitals of the electrons can also be disordered[30] since there are two degenerate orbitals for the $e_g$ electron on each Mn site.

The presence of oxygen vacancies in the ultrathin films can be considered as another important type of disorder, which is observed and evidenced by comparing the oxygen content in the 8 u.c. and 50 u.c. films using an aberration-corrected annular bright field (ABF) imaging technique. Here, the 50 u.c. film can be regarded as a reference sample since it displays bulk-like properties and is believed to possess few disorders. Since the contrast in the ABF micrograph is approximately proportional to $Z^{1/3}$, oxygen content can be detected at atomic scale[6]. The cross-sectional ABF micrographs of the LSMO film with thickness of 8 u.c. and 50 u.c. is displayed in Fig. 4a and b, respectively, while Fig. 4c and d show the corresponding line profiles of the three $MnO_2$ layers (the dark cyan bars) indicated by Ⅰ, Ⅲ and Ⅴ in Fig. 4a and b, respectively. Here, layer Ⅰ, Ⅲ and Ⅴ corresponds to the first, third and fifth $MnO_2$ layer from the interface to the surface and the oxygen content is proportional to the depth of the valley indicated by the red arrows in the line profiles. The line profiles of the ABF images in Fig. 4c and d clearly show a deficiency of oxygen content in the 8 u.c. film compared with that of the 50 u.c. film, which indicates the existence of oxygen vacancies in the 8 u.c. film.

Since oxygen vacancies can result in the decrease of the Mn valence[6,31,32], the existence of oxygen vacancies in the 8 u.c. film can be further confirmed by the electron energy loss spectroscopy (EELS). The O-K edges acquired from the 8 u.c. and 50 u.c. films are shown in Fig. 4e. The O-K edge fine structure contains information on excitations from O 1s electrons to 2p bands[33] and the three main structure peaks are indicated by A, B and C. Peak A has a strong association with the filling of the Mn 3d bands and the intensity of peak A has been found to decrease with a lowering of Mn valence[31]. We observe from Fig. 4e that peak A is almost not present in the 8 u.c. film, which indicates the decrease of the Mn valence as well as the existence of the oxygen vacancies in the 8 u.c. film. In addition, it is worth to mention that our films are deposited under the same oxygen pressure of 30 Pa and similar phenomenon of oxygen deficiency has been found in the 6 u.c. and 10 u.c. films as well. This interesting phenomenon indicates that the existence of the oxygen vacancies in ultrathin films is mainly due to the reduction in dimensionality, and further studies are still needed to understand this phenomenon of oxygen deficiency in ultrathin films.

In order to better understand the effect of oxygen vacancies on the Mn valence of the 8 u.c. film, the Mn 2p and Mn 3s X-ray photoelectron spectra (XPS) of the 8 u.c. and 50 u.c. LSMO films were obtained as well, as shown in Fig. 5a-c. It can be clearly seen from Fig. 5a that the Mn $2p_{3/2}$ binding energy was shifted from 641.3 eV to 640.7eV when the thickness of LSMO decreases from 50 u.c. to 8 u.c., indicating a decrease of the Mn valence state due to the existence of oxygen vacancies in the 8 u.c. film. To further analyze the exact Mn oxidation state in the LSMO films, we used Mn 3s exchange splitting energy $\Delta E_{3s}$ which has a linear dependence on the Mn oxidation state, and is more sensitive to oxygen defects and increases with oxygen vacancy concentration[34]. Thus we have measured $\Delta E_{3s}$ of the 8 u.c. and 50 u.c. LSMO film to estimate the Mn valence of the films using the equation[34]

$$V_{Mn} = 9.67 - \frac{1.27 \Delta E_{3s}}{eV}.$$

Fig. 5b shows that $\Delta E_{3s}$ of the 8 u.c. film is 5.16 eV which gives a Mn valence of 3.12, while $\Delta E_{3s}$ of the 50 u.c. film is 5.10 eV (Fig. 5c) and the corresponding Mn valence is 3.2.

Therefore, for the 50 u.c. film, both the Mn valence and the saturation magnetization exhibits bulk-like values, while the lowered Mn valence in the 8 u.c. film indicates the decrease of the effective doping density[12] and the reduction of the $Mn^{4+}$ content in the film, leading to the suppression of the double-exchange mechanism. Since both the magnetic and electric properties are dominated by the completion of the double-exchange and super-exchange mechanism[35], this decrease of Mn valence results in the decrease of the Curie temperature *Tc* and the suppression of the magnetic properties of the ultrathin films.

Actually, this decrease of Mn valence is in accordance with the results we obtained from our magnetic measurements, which shows suppression of magnetic properties in ultrathin films. As can be seen in Fig. 2a, in ultrathin films with resistivity upturn, the magnetic moments are smaller than those in the thicker films, indicating that the magnetic moments are not all aligned in the same direction. In fact, this disorder of magnetic moment induced in the LSMO ultrathin films should be considered as a most important disorder effect. Due to the Hund's rule coupling effect, the conduction electron of one spin will feel a high potential where the magnetic moment is along a different orientation. Therefore, the disorder of the magnetic moments is larger and the wave functions of the electrons become less extended, i.e. the localization lengths are smaller, which can be the reason why the resistivity upturn can emerge in ultrathin films.

**Summary**

In conclusion, we systematically studied the insulating phase at low temperature appeared in the LSMO films with thicknesses varying from 6 u.c. to 10 u.c. Through analyzing the transport data, we have found that the low temperature insulating behaviors of these ultrathin films can be well described by the Mott VRH law $\rho = \rho_0 \exp\left(T_0 / T\right)^{0.25}$ with a decrease of the localization length $\xi_L$ in thinner films. Further reducing of the film thickness causes an ES type VRH behavior. Meanwhile, oxygen deficiency and the resulting decrease of the Mn valence have been observed in ultrathin films, and we proved that this existence of oxygen vacancies due to the reduced dimensionality can reduce the effective doping density and the $Mn^{4+}/Mn^{3+}$ ratio, thus suppresses the double-exchange mechanism in ultrathin films. Therefore, both the magnetic and transport properties of the ultrathin films are suppressed. In addition, the suppression of the magnetic properties in thinner films indicates stronger disorder of magnetic moments, which is considered to be the reason for this decrease of the localization length $\xi_L$. These results suggest that the concentration of the oxygen vacancy in thinner films is one reason for the enhancement of disorder, which leads to the carrier localization and the resistivity upturn.

## Methods

**Sample preparation**. LSMO films were deposited on (001) cut $TiO_2$-terminated STO substrates by a computer-controlled laser molecular beam epitaxy (Laser-MBE) equipped with an *in situ* reflection high-energy electron diffraction (RHEED) system (PASCAL) allowing for precise control of the thickness at the atomic scale. A XeCl 308 nm excimer laser was used with an energy density of 2.18 J/cm$^2$ and a repetition rate of 2 Hz. The films were deposited at 930 ℃ with the oxygen pressure of 30 Pa. After deposition, the samples were in-situ annealed for 10 min, and then cooled down to room temperature. The surface morphology of the LSMO films was recorded using a commercial atomic force microscopy (AFM) system (Asylum Research MFP3D).

**Sample characterization.** The crystal structure was identified by high-resolution Synchrotron X-ray diffractometry（XRD） by the BL14B1 beam line of Shanghai Synchrotron Radiation Facility (SSRF), using a 1.24 Å X-rays with a Huber 5021 six-axis diffractometry and the thicknesses of the films which were measured by X-ray reflectometry（XRR）agree well with the number of the RHEED oscillations during deposition. Both the transport and magnetic properties of the thin films were measured with a Quantum Design physical property measurement system (PPMS).

The transport properties of the thin films were measured using a four-probe method in the temperature range from 10 to 400 K. The atomic structure of the LSMO thin films was characterized using an ARM-200CF (JEOL, Tokyo, Japan) transmission electron microscope (TEM) operated at 200 keV and equipped with double spherical aberration (Cs) correctors. The attainable resolution of the probe defined by the objective pre-field is 78 picometers. Electron Energy Loss Spectroscopy (EELS) were acquired with a multi-scan charge-coupled device (CCD) camera (Gatan Quantum Model 965, Gatan Inc.). X-ray photoelectron spectroscopy (XPS) is used to obtain the electronic structure and chemical states of the LSMO films at the surface and the measurement temperature for XPS is 300 K.

**Acknowledgement**

This work is Supported by the "Strategic Priority Research Program（B）" of the Chinese Academy of Sciences，Grant No. XDB07030200, and the National Natural Science Foundation of China (Nos. 11134012, 11404380). The XRR and XRD measurements were supported by the Shanghai Synchrotron Radiation Facility (SSRF).

**Author contributions**

Y-q. Feng and K-j. Jin contributed the whole idea and designed the experiments. Y-q. Feng and Meng. He prepared the films and devices. Y-q. Feng and C. Ge measured the electrical and magnetic properties of the films and conducted the AFM experiments. X. He performed the theoretical analysis. Q. Wan performed the XRD test and analysis. Q-h. Zhang, L. Gu and Min. He helped to collect the TEM images and analyze the EELS data. Q-l. Guo performed the XPS test and analysis. Y-q. Feng, K-j. Jin, C. Ge and X. He wrote the paper. All the authors discussed the results and commented on the manuscript.

**Additional Information**

**Competing financial interests:** The authors declare no competing financial interests.

**Figure legends**

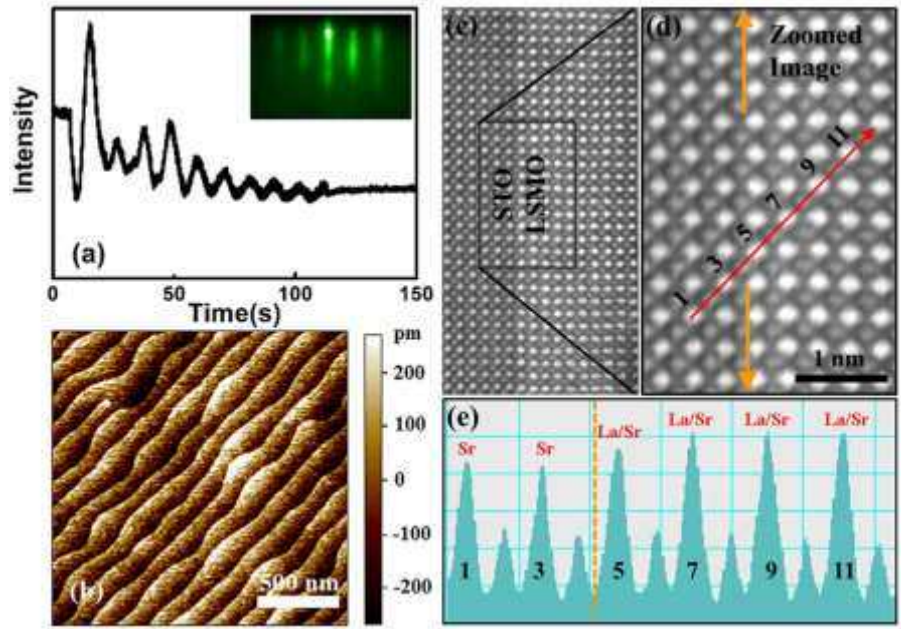

**Figure 1 | Samples preparation and characterization.** (**a**) RHEED oscillations recorded during the growth of the 10 u.c. LSMO film and (**b**) AFM image of the 10 u.c. film. (**c**) HAADF micrograph of the 10 u.c. LSMO thin film taken by an aberration-corrected STEM. (**d**) Zoomed and filtered image of the rectangular part in (**c**) with the red arrow of numbered peaks. The yellow arrows indicate where the STO/LSMO interface locates. (**e**) Intensity line profile of the region marked by the red arrow in (**d**). The different contrast of La and Sr provides localization of the interface indicated by the yellow dashed line in (**e**).

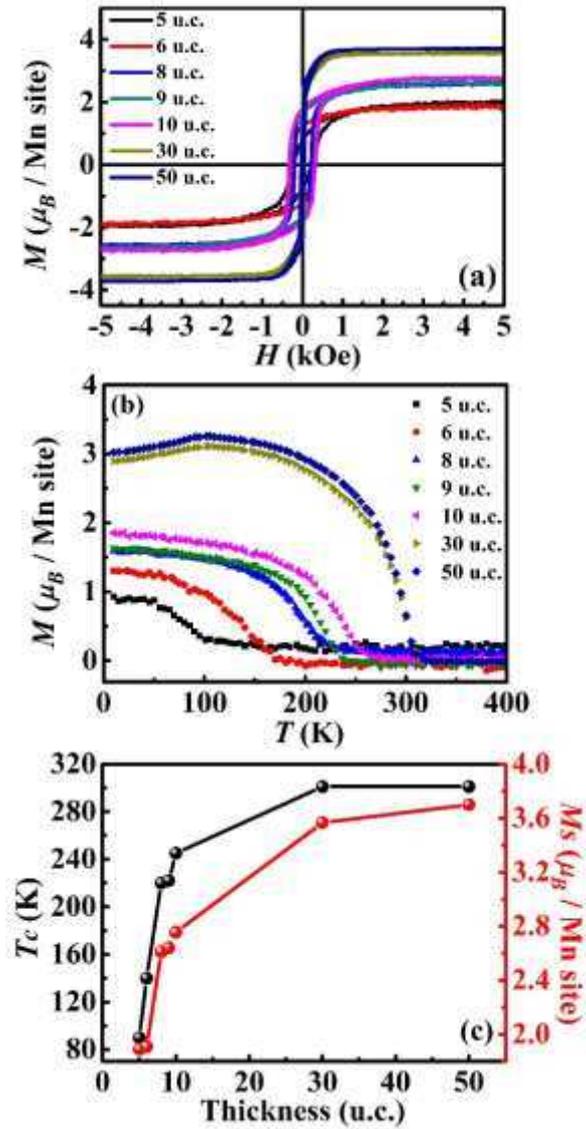

**Figure 2 | Magnetic properties of LSMO films with different thickness.** (**a**) The hysteresis loops (10 K) of the LSMO films and (**b**) the temperature dependence of magnetization (200 Oe, field cooling) of the LSMO films. (**c**) The thickness-dependent saturation magnetization *Ms* and Curie temperature *Tc* of the LSMO films.

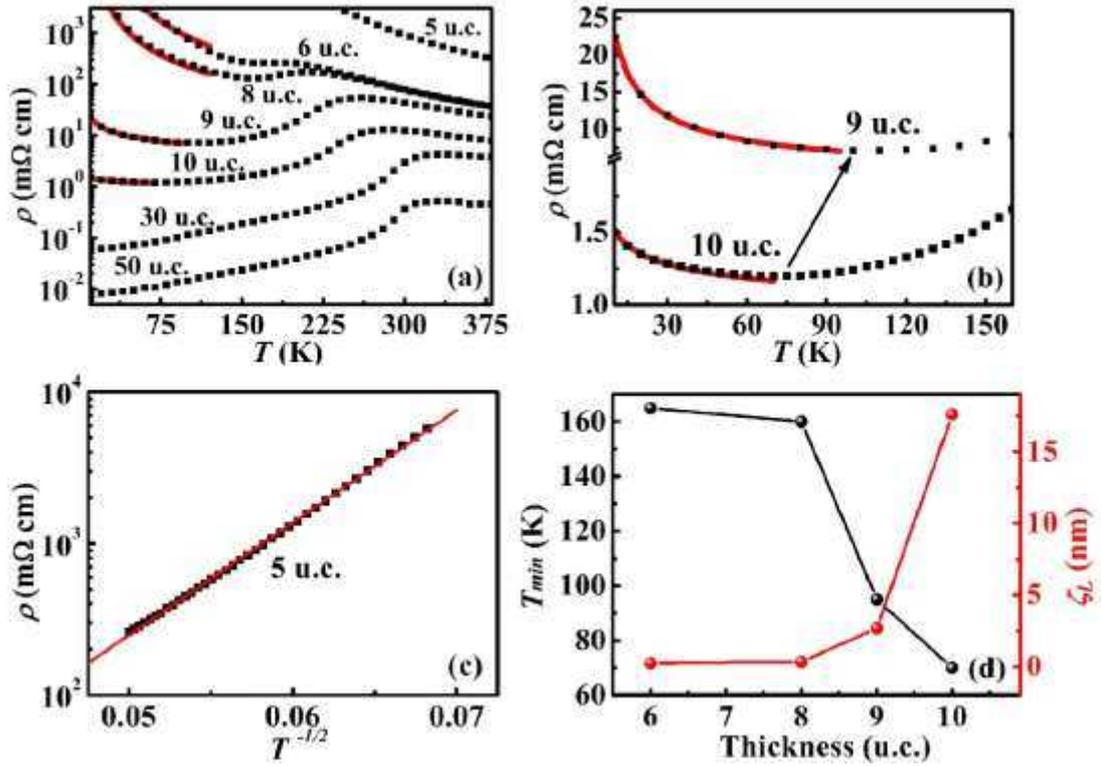

**Figure 3 | Transport properties of LSMO films with different thickness.** (**a**) Temperature dependent resistivity of LSMO films with various thicknesses. (**b**) Temperature dependent resistivity of the 9 u.c. and 10 u.c. films, the arrow indicates variation of the resistivity upturn between the two films. The red solid lines in (**a**) and (**b**) show the fitting results of the low temperature insulating behaviors of the films with resistivity upturn by the 3D Mott VRH mechanism. (**c**) Log plot of resistivity as a function of $T^{-1/2}$ for the 5 u.c. film. (**d**) The thickness-dependent *Tmin* and the localization lengths $\xi_L$ of the ultrathin films with resistivity upturn.

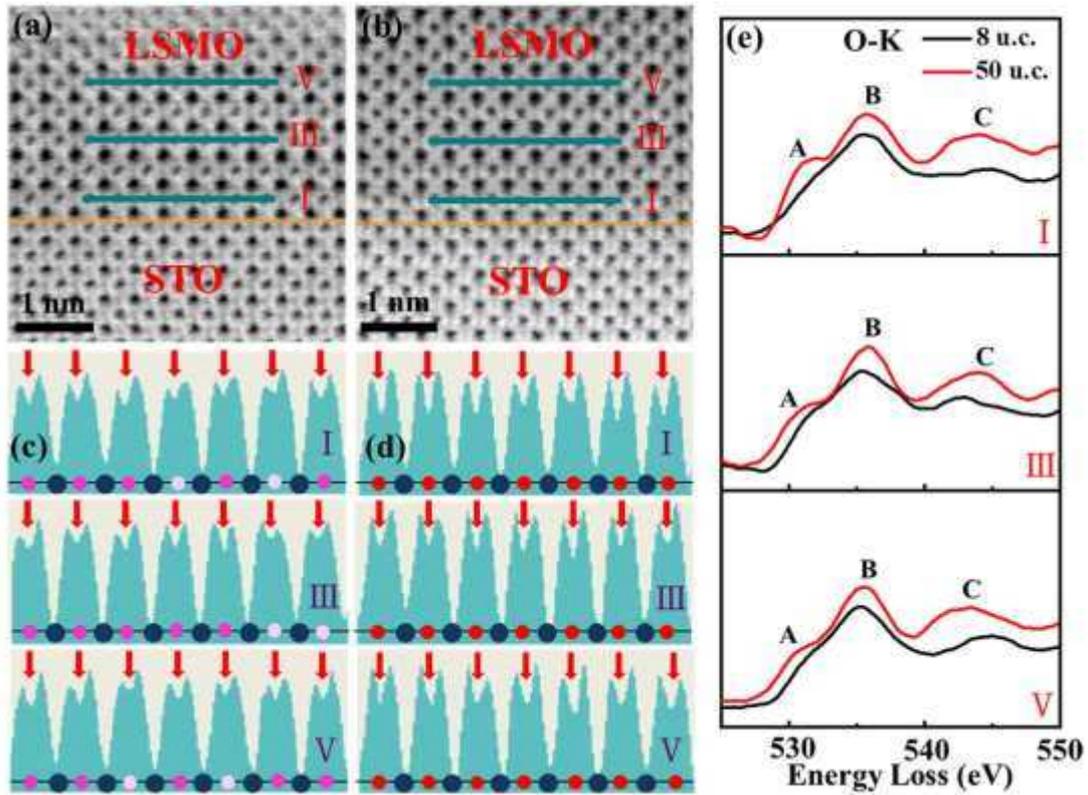

**Figure 4 | Comparative study of oxygen content in the 50 u.c. and 8 u.c. films.** (**a**) The cross sectional ABF micrographs of the 8 u.c. and (**b**) 50 u.c. LSMO films taken by an aberration-corrected STEM. (**c**)- (**d**) The line profiles of the three $MnO_2$ layers (indicated by Ⅰ, Ⅲ and Ⅴ in the ABF images). The yellow dashed lines indicate where the interface locates. The red arrows point at the O sites. The blue and red circles indicate Mn and O ions, respectively. The lighter red and white circle indicates the presence of oxygen vacancies. (**e**) Comparison of EEL spectra of the O-K edge acquired from the 8 u.c. and 50 u.c. LSMO films. Ⅰ, Ⅲ and Ⅴ corresponds to the three $MnO_2$ layers marked in (**a**) and (**b**).

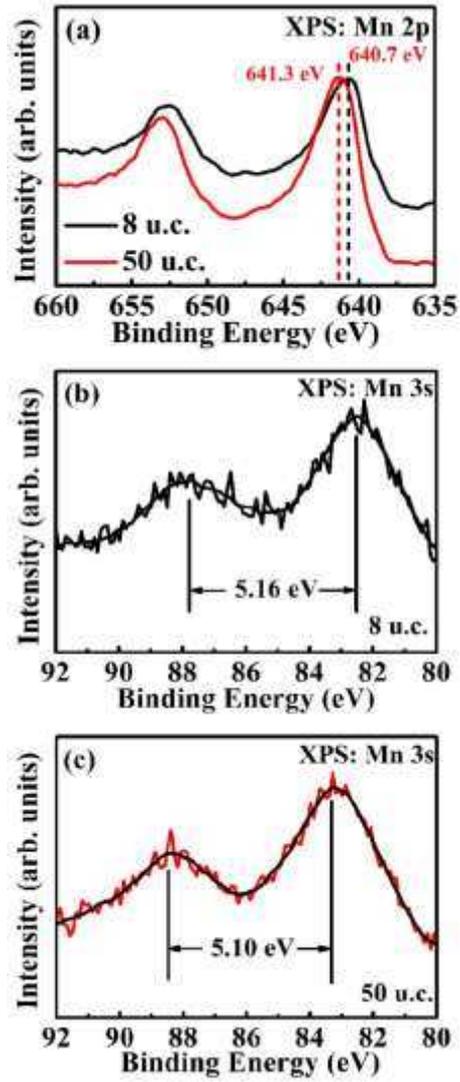

**Figure 5 | X-ray photoelectron spectra of the 8 u.c. and 50 u.c. films.** (**a**) Mn 2p XP spectra of the 8 u.c. and 50 u.c. LSMO films, respectively. (**b**) Mn 3s XP spectra of the 8 u.c. and (**c**) 50 u.c. LSMO films.

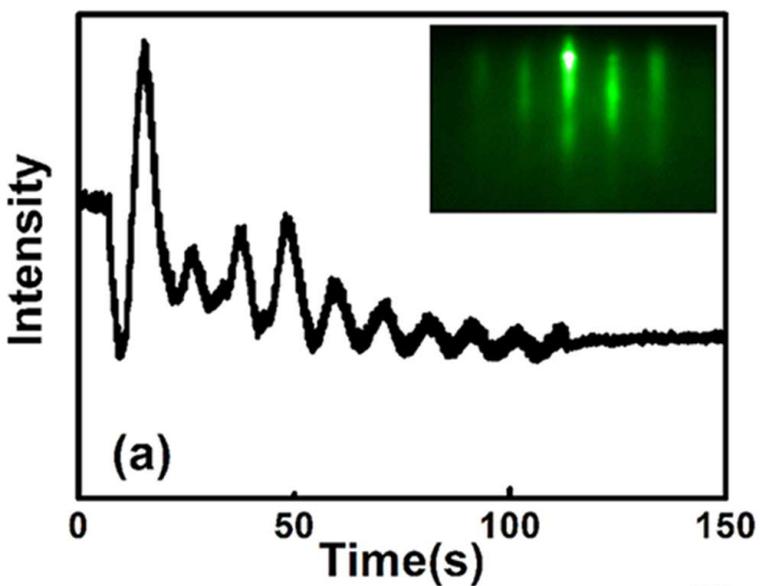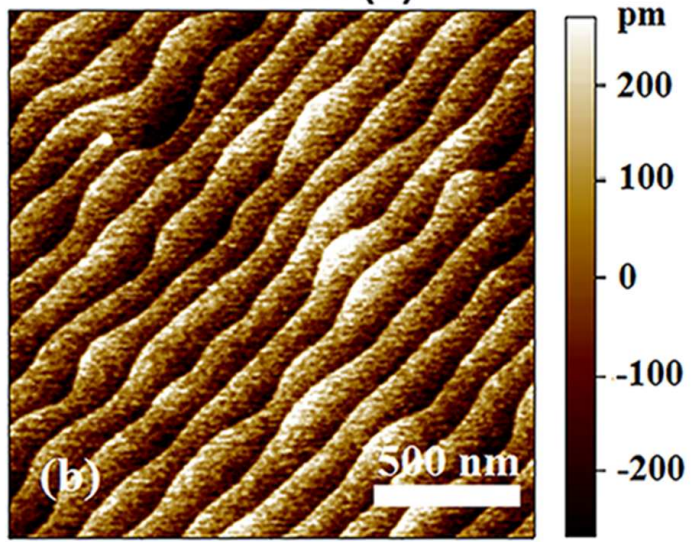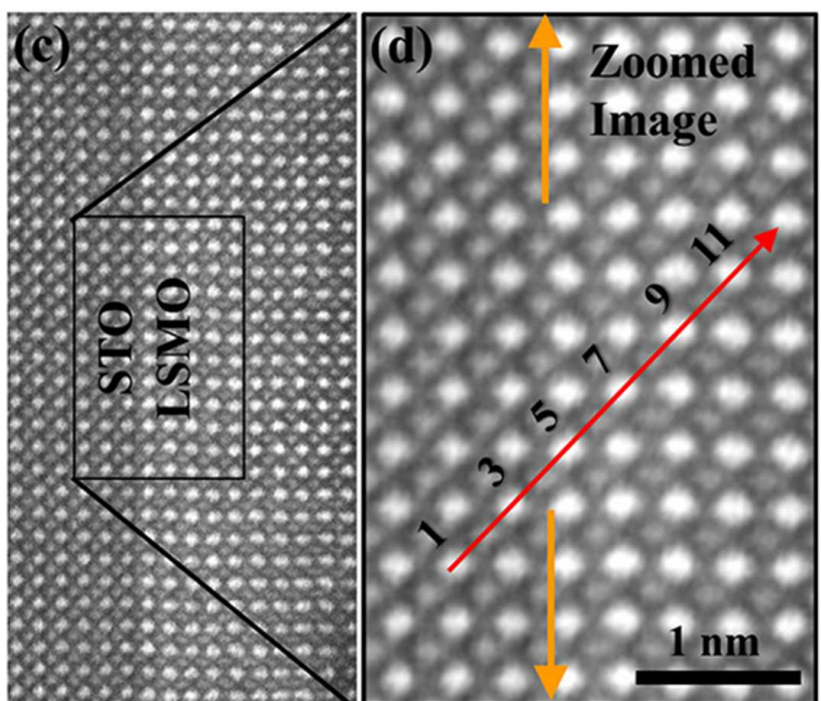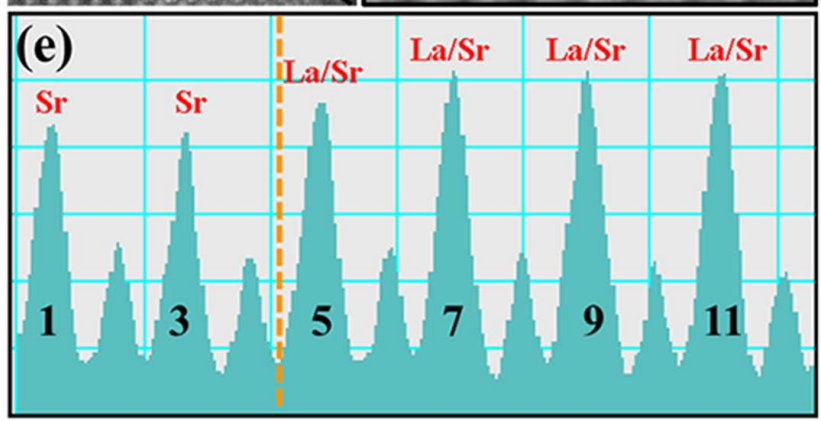

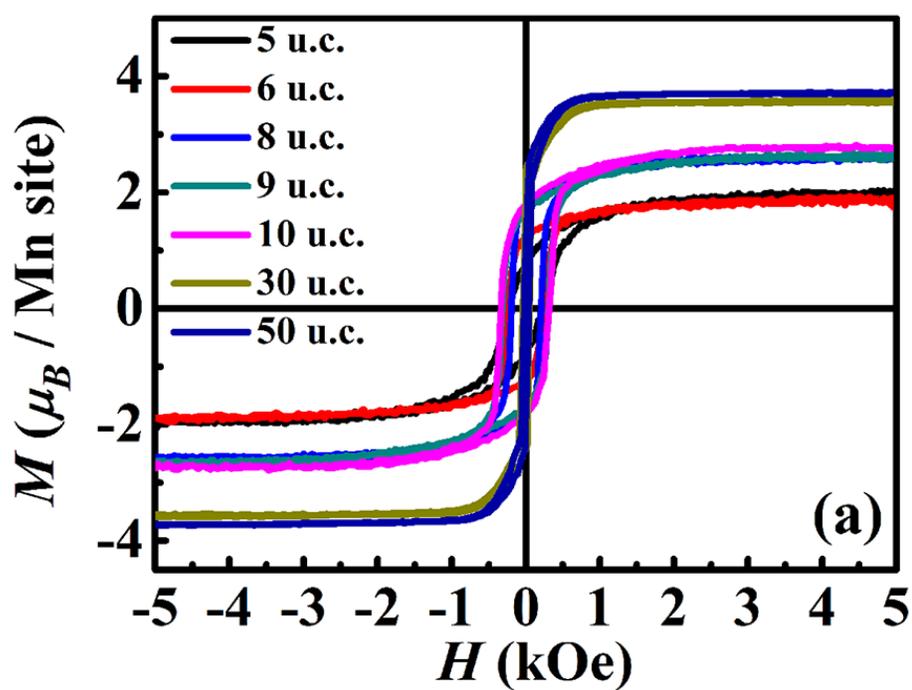
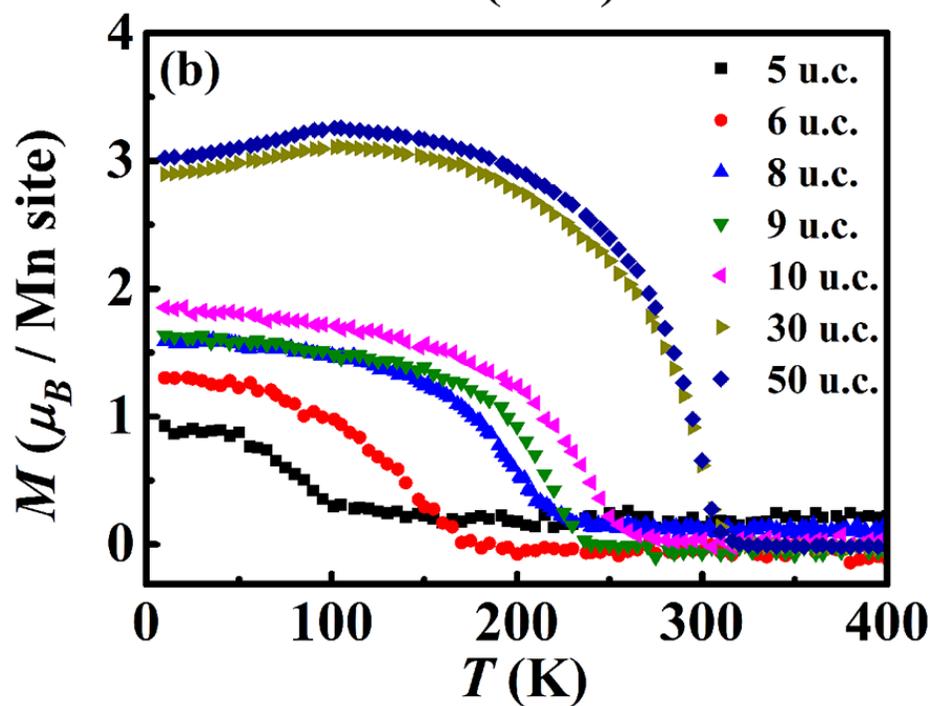
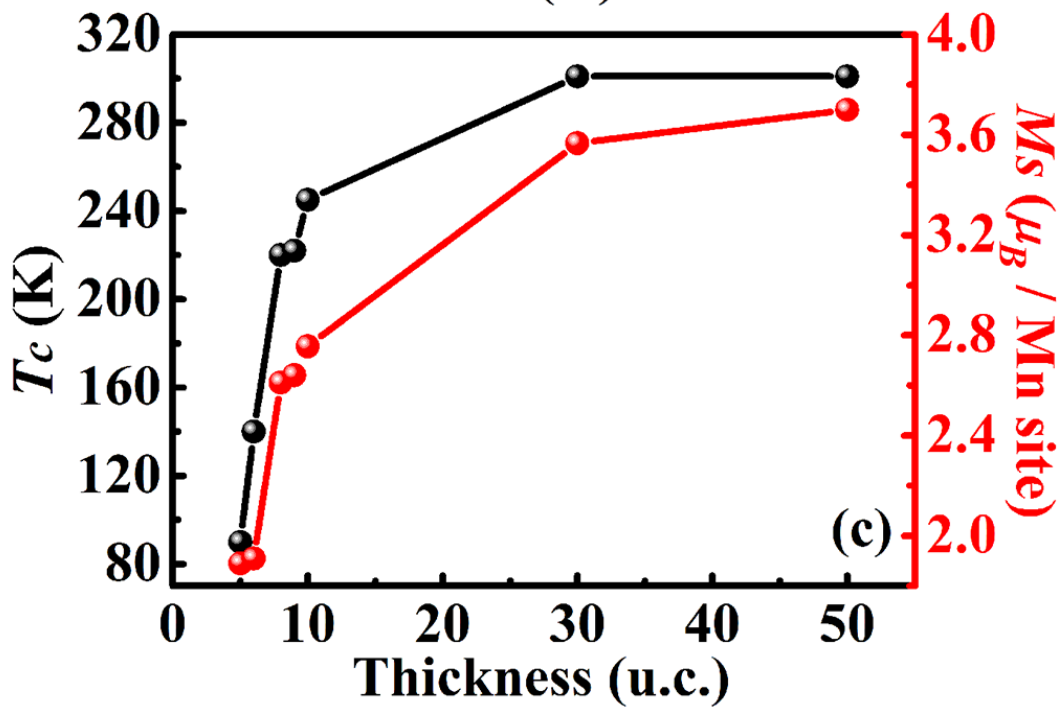

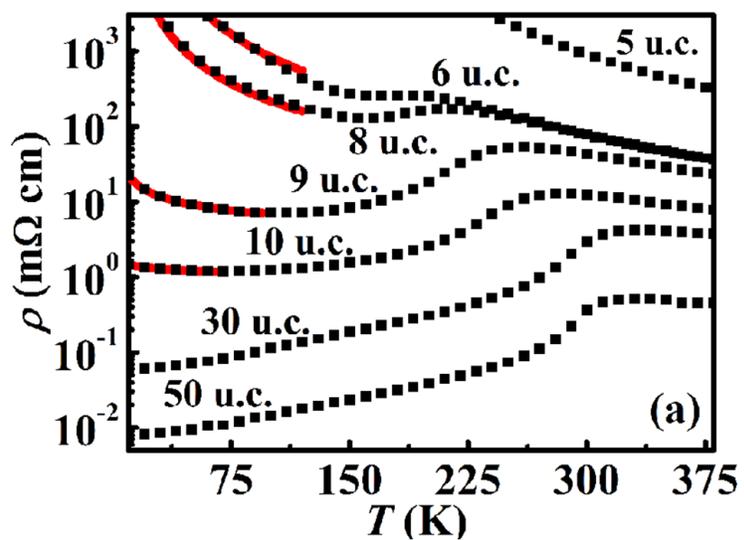
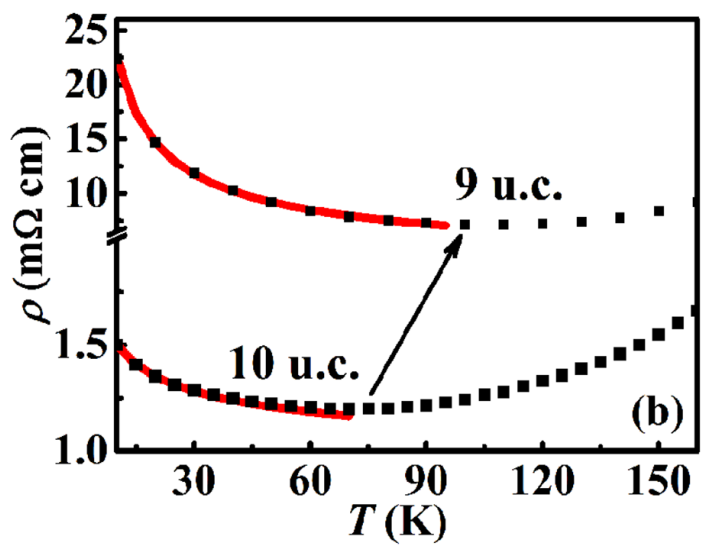
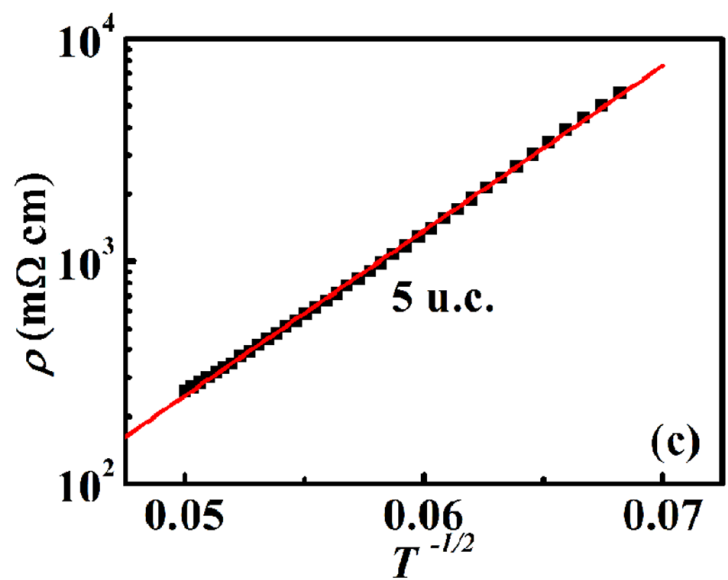
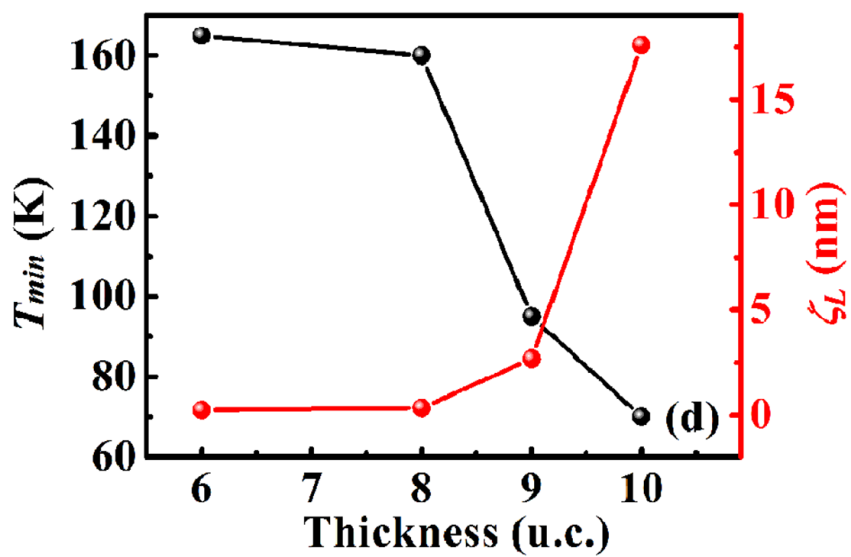

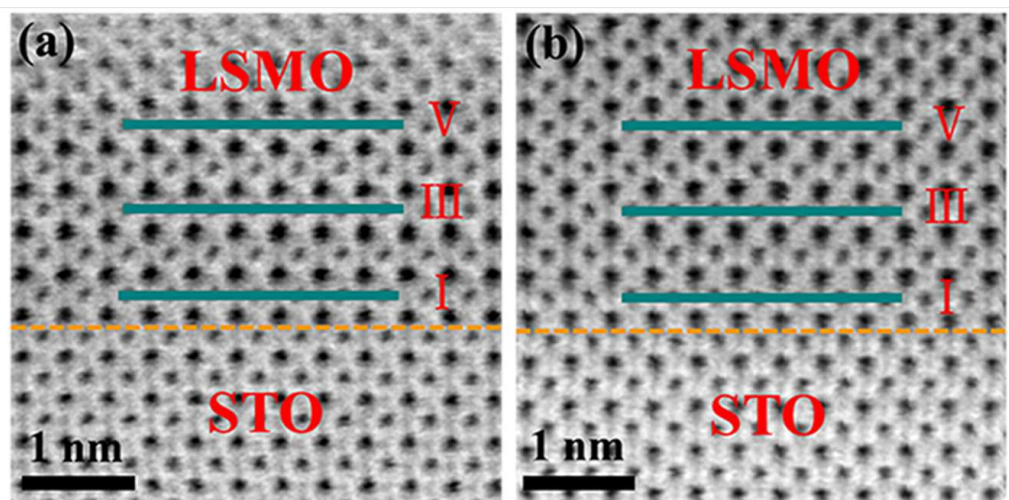
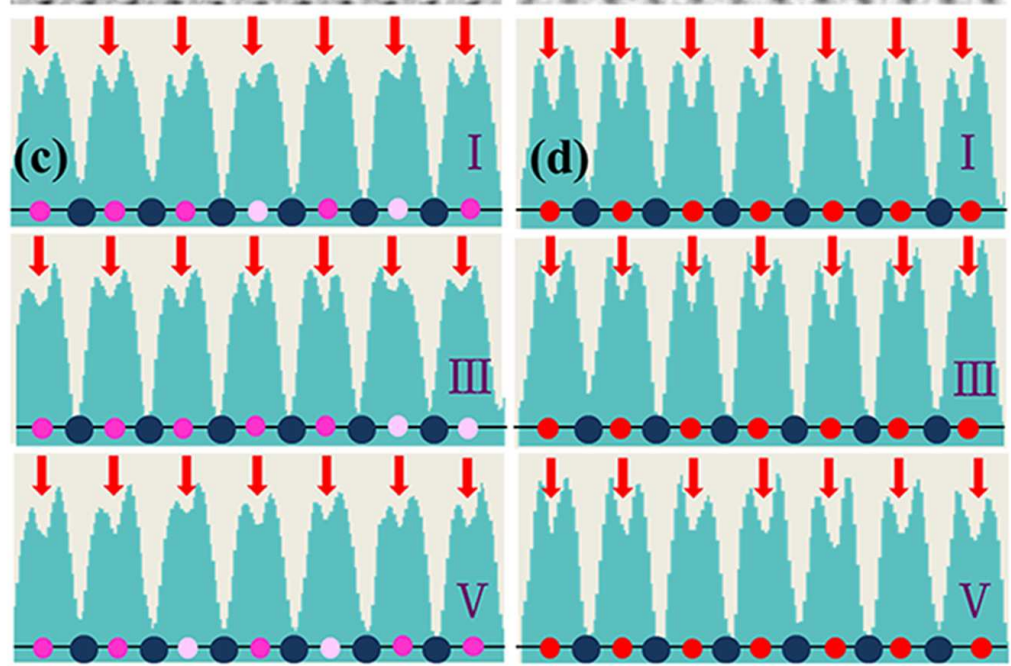
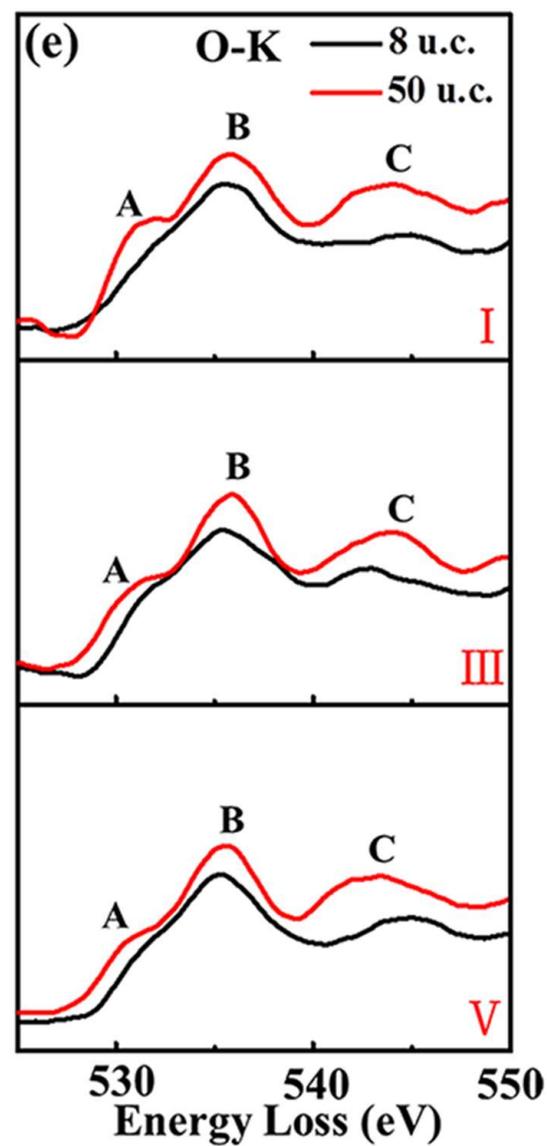

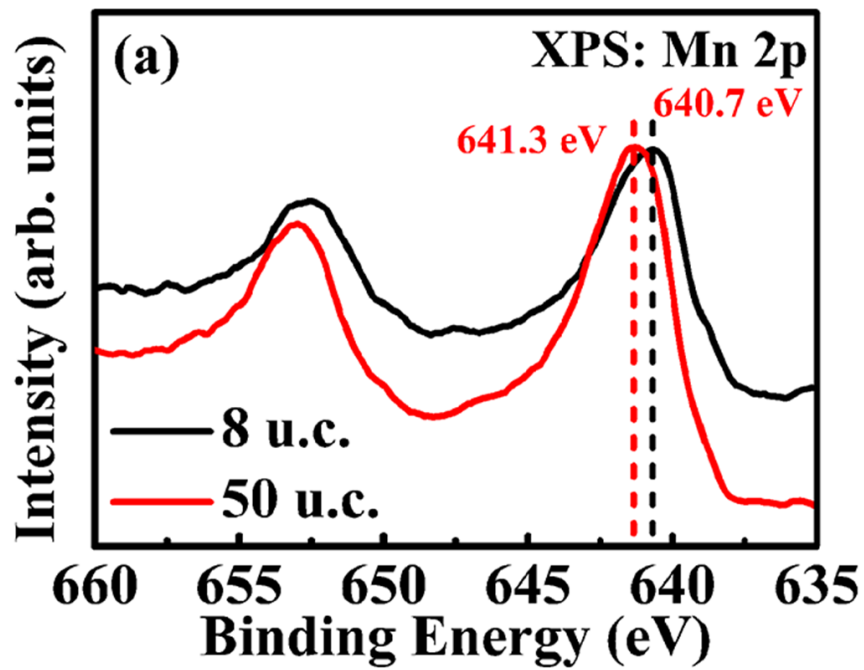
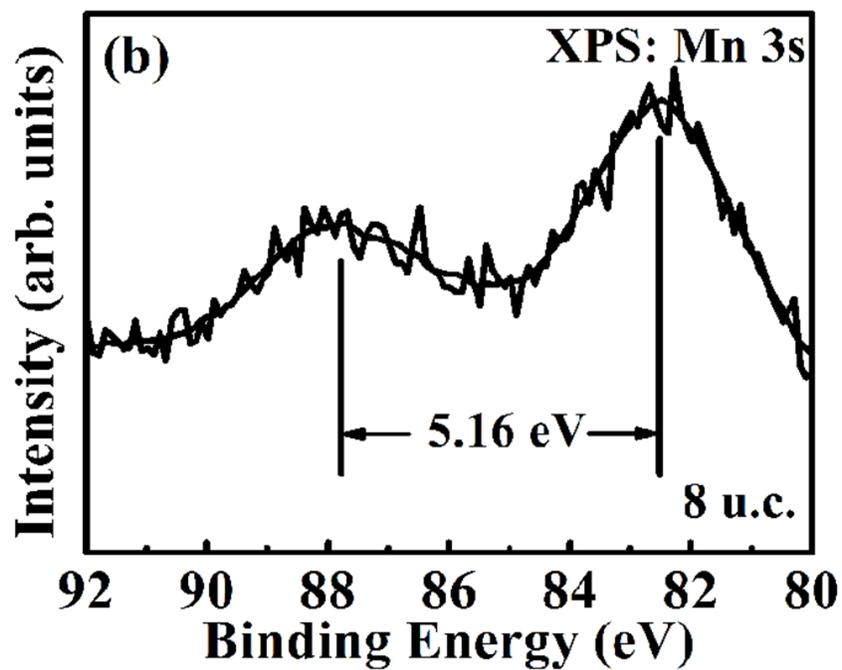
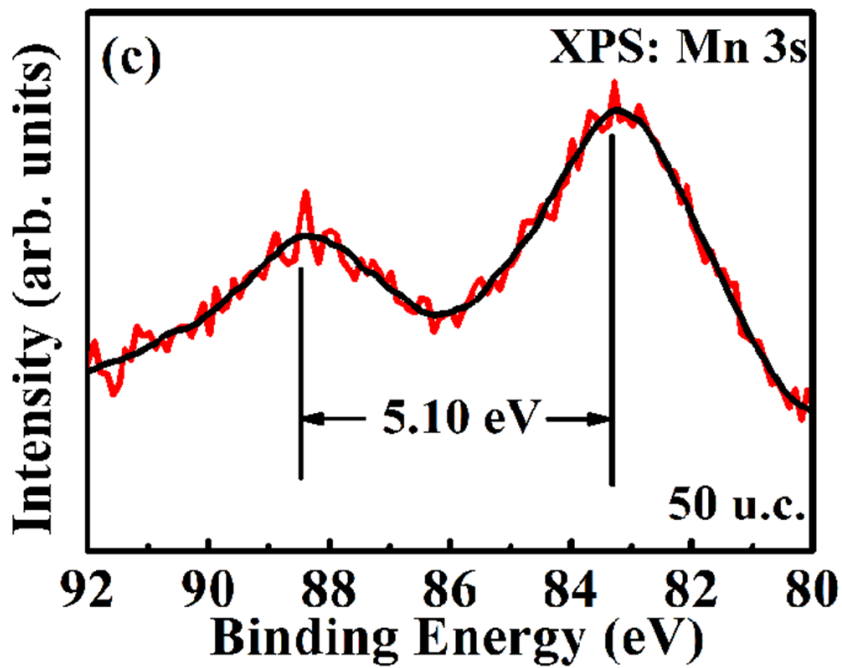